# Radial thermal expansion of pure and Xe-saturated bundles of single-walled carbon nanotubes at low temperatures

A.V. Dolbin[1], V.B. Esel'son[1], V.G. Gavrilko[1], V.G. Manzhelii[1], S.N. Popov[1], N.A.Vinnikov[1], N.I. Danilenko[2], B. Sundqvist[3].

[1] B. Verkin Institute for Low Temperature Physics and Engineering of the National Academy of Sciences of Ukraine, 47 Lenin Ave., Kharkov 61103, Ukraine
[2] Frantsevich Institute for Problems of Materials Science of the National Academy of Sciences of Ukraine, Krzhizhanovsky str., 3, Kyiv 03680, Ukraine
[3] Department of Physics, Umea University, SE - 901 87 Umea, Sweden

Electronic address: dolbin@ilt.kharkov.ua

## Abstract

The radial thermal expansion coefficient $\alpha_r$ of pure and Xe-saturated bundles of single-walled carbon nanotubes has been measured in the interval 2.2—120 K. The coefficient is positive above T = 5.5 K and negative at lower temperatures. The experiment was made using a low temperature capacitance dilatometer with a sensitivity of $2 \cdot 10^{-9}$ cm and the sample was prepared by compacting a CNT powder such that the pressure applied oriented the nanotube axes perpendicular to the axis of the cylindrical sample. The data show that individual nanotubes have a negative thermal expansion while the solid compacted material has a positive expansion coefficient due to expansion of the intertube volume in the bundles. Doping the nanotubes with Xe caused a sharp increase in the magnitude of $\alpha_r$ in the whole range of temperatures used, and a peak in the dependence $\alpha_r (T)$ in the interval 50—65 K. A subsequent decrease in the Xe concentration lowered the peak considerably but had little effect on the thermal expansion coefficient of the sample outside the region of the peak. The features revealed have been explained qualitatively.

## Introduction

Since the discovery of carbon nanotubes (CNTs) in 1991 [1], this novel class of physical objects has been stimulating intense experimental and theoretical research activities. The diversity of CNT types and the problems encountered in obtaining pure CNT material in quantities needed for experimental investigations make it rather difficult to trace the basic trends in the behavior of carbon nanotubes (e.g., see the text and References in [2]). Thermal expansion is one of the least studied properties of CNTs. The currently available experimental evidence on the thermal expansion of single-walled nanotubes (SWNTs) and their bundles is confined to the region near and above room temperature, whereas low temperature data are essential for understanding the CNT dynamics. The theoretically estimated thermal expansion coefficients (TEC) of SWNTs [3—9] vary appreciably both in magnitude and sign.

Owing to their unique geometry, CNTs can be a basis for forming novel low-dimensional systems. For example, bundles can be used as templates to form one-dimensional chains or two-dimensional surfaces consisting of condensed impurity molecules.

In recent years much experimental effort has been devoted to the study of structural and thermal properties of such systems and a number of theoretical models have been advanced to predict these properties [10—23]. However, the thermal expansion behavior of SWNT—gas impurity systems still remains obscure.

In this study the radial thermal expansion was measured on a sample consisting of bundles of single-walled nanotubes closed at the ends (c-SWNT) in the range T = 2.2—120 K and on bundles of SWNTs saturated with Xe at T = 2.2—75 K. The sorption properties of bundles of SWNTs with closed (c-SWNT) and open (o-SWNT) ends were investigated using the technique described below.

# 1.Radial thermal expansion of pure single-walled carbon nanotubes

## *1.1.Measurement technique and investigated sample*

The sample for thermal expansion measurements was prepared using a procedure for ordering the SWNT axes by applying a pressure of 1 GPa, as described by Bendiab et al. [24]. These authors showed that in SWNT plates of up to 0.4 mm thickness, such a pressure aligned the CNT axes in the sample such that their average angular deviation from a plane normal to the pressure vector was ~4º.

The starting material was a CNT powder (CCVD method, Cheap Tubes, USA) which according to the manufacturer contained over 90% of SWNTs. The main characteristics are given in the Table.

Table I. Characteristics of carbon nanotube powder given by the manufacturer.

| Diameter | l-2 nm |
|---|---|
| Length | 5-30 μm |
| SWNT fraction | > 90 wt % |
| Amorphous carbon fraction | < 1.5 wt % |
| Co catalyst fraction | 2,9 wt % |
| Specific surface | > 407 $m^2$/g |
| Electrical conduction | > 102 S/cm |

The quality of the powder was confirmed by Raman analysis performed both by the supplying company and at Umeå University, Sweden. According to the manufacturer, the average outer diameter of the tubes was 1.1 nm but no information is available about the chirality distribution. From our own Raman data, obtained using four different excitation lasers with wavelengths in the range 541-830 nm, we find that the radial breathing modes indicate a wide range of tube diameters, 0.8 – 2.1 nm. All samples studied show typical SWNT G-bands and only weak disorder bands. Although multi-wall tubes may also be present, judging from the spread in diameters, the Raman spectra are completely dominated by the response from single (or possibly few-) wall nanotubes. However, a small fraction of MWNTs might be invisible due to their large diameters and possibly lower Raman cross sections.

The starting SWNT powder was also investigated by high-resolution transmission electron microscopy (HRTEM) at both the Institute of Problems of Material Science, NAS of Ukraine (Fig. 1a) and at Umea University, Sweden (Fig. 1b). The pictures show that large sample fractions contain little amorphous carbon or residual catalyst. By measuring the bundle diameters we estimate that in the starting powder each bundle contains 7 to 600 SWNTs.

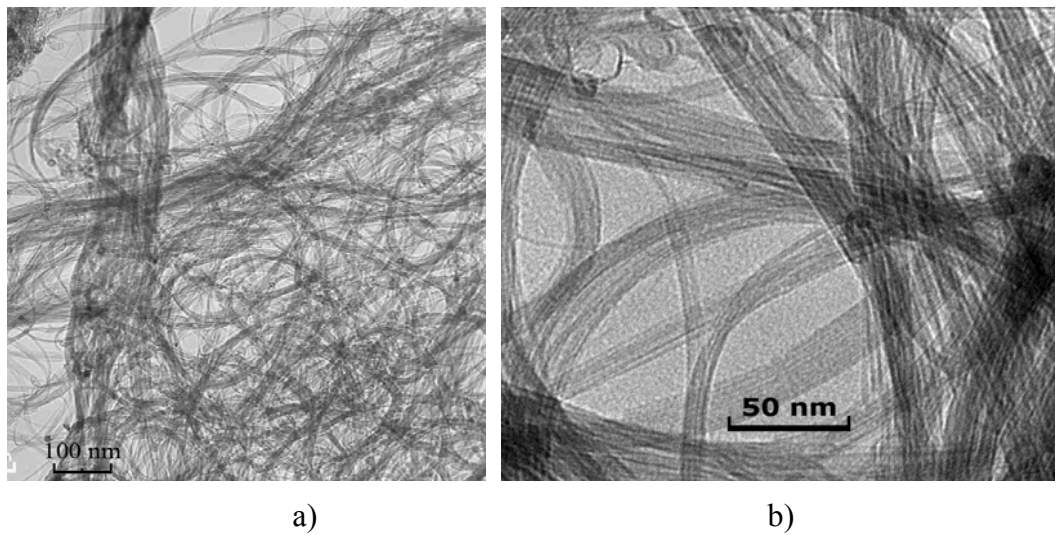

a)                               b)

Fig. 1 TEM images of the starting SWNT powder.

The compacted sample used was prepared at Umea University (Sweden) by first compacting pressure-oriented (P = 1.1 GPa) SWNT plates (an individual plate was up to 0.4 mm thick), then pressing several stacked plates together at a ten percent higher pressure to form a cylinder 7.2 mm high and 10 mm in diameter with a density of 1.2 g/cm$^3$. The sample was made in a special cylindrical segmented die designed for compacting CNT powder under effective pressures 0.5—2 GPa, consisting of a ring with a cylindrical channel and a conical outer surface, which was inserted into a hardened-steel cylinder supported inside a larger pressure vessel. The structure so arranged was resistant to internal stresses. The 10 mm in diameter piston was made from sintered tungsten carbide (WC). The pressures used were high enough to consolidate the powder into a solid with well oriented tubes [24], but still low enough to keep the integrity and structure of the tubes and avoid tube collapse, and Raman spectra taken on pressed plates showed no systematic changes relative to spectra taken on the pristine powder. The sample prepared by this technology has a pronounced anisotropy of properties in the directions perpendicular and parallel to the sample axis. In the direction perpendicular to the applied pressure the axes of the SWNT bundles are disordered. The compaction aligns the axes of the SWNT bundles in the plane perpendicular to the sample axis [24]. As a result the radial component of the expansion of the SWNT bundles makes a dominant contribution to the thermal expansion of the sample in the directional parallel to the sample axis. If the axial component of the thermal expansion coefficient has a magnitude comparable to that of the radial one, an angle of typically 4$^o$ implies that the typical contribution to the total coefficient from the axial component is about 7 % of the magnitude of the radial component.

The radial thermal expansion of the sample was investigated using a capacitance dilatometer (its design and the measurement technique are described in [25]). The linear thermal expansion coefficient (LTEC) was measured in the direction of the applied compacting pressure, i.e. radially to the SWNT bundles. Prior to measurement, the gas impurities were removed from the sample by dynamic evacuation for 72 hours at 10$^{-3}$ mm Hg and room temperature. Immediately before dilatometric investigation, the measuring cell with the sample was cooled slowly (for 8 hours) down to liquid helium temperature (4.2 K) and the sample was held at this temperature for about 4 hours. The cooling and investigation were made in vacuum down to 10$^{-5}$ mm Hg.

## 1.2. Experimental results and discussion

The temperature dependence of the LTEC in the interval 2.2—120 K is shown in Fig. 2. The curves were obtained by least-square averaging over several series of measurement.

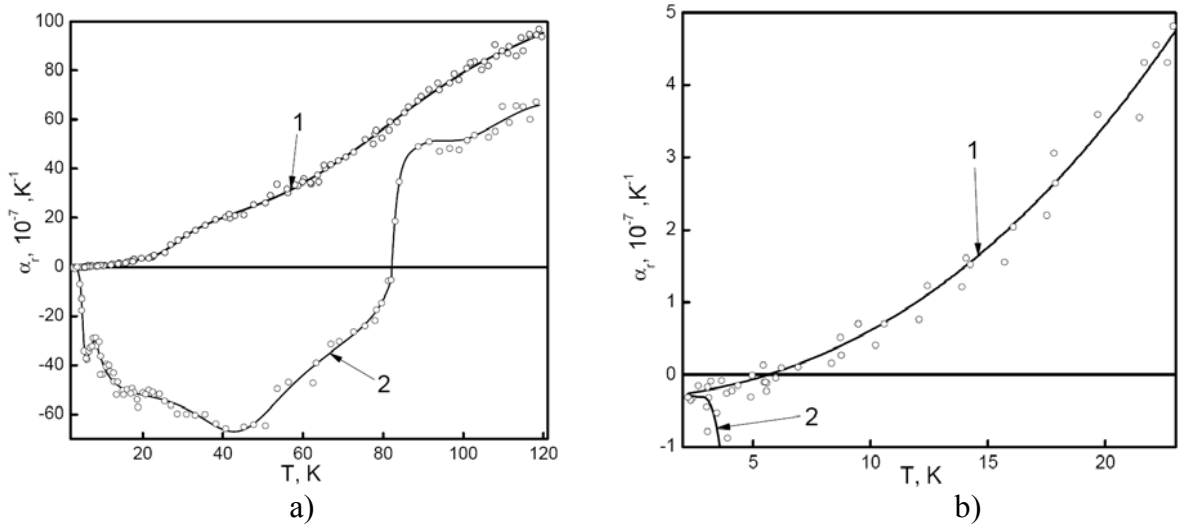

Fig. 2. LTECs of pressure-oriented SWNT compacted sample in the direction perpendicular to the SWNT bundle axes: a) T = 2.2—120 K; b) T = 2.2—25 K (curve 1 — heating and cooling, curve 2 — first heating from T = 2.2 K).

Curve 2 was taken on the first heating of the sample from T = 2.2 K. Curve 1 data were measured in the subsequent heating-cooling process. The non-equilibrium LTECs obtained on the first heating from T = 2.2 K may account for heating-induced alignment and ordering of the bundles and the nanotubes in them, which causes a compression of bundles and, as a result, negative thermal expansion.

The equilibrium radial LTEC $\alpha_r$ (curve 1) is positive above 5.5 K and negative at lower temperatures.

Assuming that the impurity effect is negligible, $\alpha_r$ comprises two components $\alpha_d$ and $\alpha_g$ accounting for temperature-induced changes in the CNT diameters and the intertube gap. From a simple Grüneisen-type model it might be expected that $\alpha_d$ should be similar to the in-plane thermal expansion of graphite, and thus probably small and negative well below room temperature. Because the sample is a mixture of all chiralities, the average $\alpha_d$ should also be very similar to the average axial expansion coefficient of the tubes. The thermal expansion of a bundle should thus probably be dominated by $\alpha_g$, which should be similar to the out-of-plane thermal expansion of graphite or, considering the curvature, to the thermal expansion of fullerenes or linear fullerene polymers [26].

So far there has been only one study [27] in which both $\alpha_d$ and $\alpha_g$ were measured by the X-ray diffraction method in the interval 300—950 K. At T = 300 K $\alpha_r = (0.75 \pm 0.25) \cdot 10^{-5}$ K$^{-1}$, $\alpha_d = (-0.15 \pm 0.2) \cdot 10^{-5}$ K$^{-1}$ and $\alpha_g = (4.2 \pm 1.4) \cdot 10^{-5}$ K$^{-1}$. Another measurement, of $\alpha_r$ only, by the same method [28] arrived at negative values in the whole range of measurement temperatures (200—1600 K). We are not aware of further experimental attempts to directly investigate the thermal expansion of SWNT bundles, but some experiments have been made to estimate the thermal expansion from the temperature dependence of the radial breathing Raman modes of nanotubes. Although these modes shift down rapidly with increasing temperature, indicating a large strong positive thermal expansion coefficient, it was concluded by Raravikar et al. [7] that this effect is almost completely caused by changes in intra- and intertube interactions, and that $\alpha_d$ is very small.

It is rather problematic to compare our results with theoretical data quantitatively, mainly because the available theoretical studies are concerned with the radial and axial thermal expansion of individual CNTs. Some of them offer general conjectures on how thermal expansion can be affected by the interaction of nanotubes in a bundle (e.g., see [8]). Also, there is little agreement between the theoretical conclusions from different groups about the TEC magnitude, sign and temperature dependence, about the effect of chirality and CNT diameter upon thermal expansion, and about the correlation between the radial and axial components of the thermal expansion of nanotubes. For example, the thermal expansion is negative in a wide temperature interval (0—800 K) in [4], changes from negative magnitudes at low temperatures to positive ones at moderate and high temperatures in [8] or is positive at all the temperatures investigated in [6].

The qualitative interpretation of our results is based on the Grüneisen coefficients calculated [8] for carbon modifications — diamond, graphene, graphite and nanotubes. It is found [8] that the Grüneisen coefficients and the radial thermal expansion of CNTs are negative at relatively low temperatures, an effect caused mainly by the contribution from transverse acoustic vibrations perpendicular to the CNT surface. However, our measurements show that a negative thermal expansion coefficient exists only in a temperature interval much more narrow than found in the calculations [8]. We believe that the main reason for this is that the calculations were performed for individual nanotubes only [8]. Our sample is clearly dominated by CNT bundles (Fig. 1), and in this case additional factors contributing to the thermal expansion come into play. Firstly, there appears a positive contribution $\alpha_g$ caused by variations of the intertube gaps with temperature. Secondly, the nanotube interaction in the bundles suppresses the negative contribution of the transverse acoustic vibrations perpendicular to the nanotube surfaces [8]. These two positive contributions to the thermal expansion of SWNT bundles decrease both the magnitude and the temperature region of the total negative thermal expansion. If we use this model and assume $\alpha_d$ to vary slowly with temperature over a wide temperature interval we can use the data shown in Fig. 2b to estimate $\alpha_d = (-4 \pm 1) \cdot 10^{-8}$ K$^{-1}$ at T = 2.2 K. Assuming further that the temperature dependent part of $\alpha$ at low temperatures is dominated by a positive coefficient $\alpha_g$, we see from Fig. 2b that a polynomial of the third order in T is a good approximation to $\alpha_g(T)$ up to about 25 K. Although the scatter in the data is somewhat high it is clear that to get a good fit it is necessary to include one term in $T^3$ and one term linear (or, with a less good fit, quadratic) in T. In a Grüneisen model, the thermal expansion coefficient of a bundle is closely related to its specific heat capacity, and it is well known that the experimentally found low-temperature specific heat of nanotube bundles shows a similar behaviour above 2 K [29]. In that case the experimental behaviour $c_p(T) = aT + bT^3$ could be fitted by an anisotropic two-band Debye model with weak coupling between tubes in the bundle by adding a contribution from the first optic branch. It thus seems quite reasonable to attribute the strongly temperature dependent positive component of the total thermal expansion to $\alpha_g$. The data in Fig. 2a also shows a noticeable plateau-like structure between 40 and 60 K. We point out that the intermolecular interaction in $C_{60}$, which should be similar in magnitude to the inter-tube interaction, corresponds to an effective Debye temperature near 50-60 K which gives rise to a plateau in the specific heat in this range for both molecular and polymeric $C_{60}$ [30]. The plateau structure observed here might thus indicate the cross-over between the acoustic modes and the lowest optical/molecular 3D modes in the bundle lattice.

## 2. Xe sorption in the powder of carbon nanotubes with closed and open ends

### *2.1. Measurement technique and investigated samples*

Carbon nanotubes (CNT) prepared by standard methods (electric-arc, laser evaporation of carbon, or CCVD method) are arranged into bundles. Inside a bundle the CNTs form a close-packed two-dimensional (2D) triangular lattice [31]. Normally, CNTs have fullerene-like semispheres at the ends (CNTs with closed ends, or c-SWNT). The final CNT product can contain large amounts of amorphous carbon, fullerenes and other carbon modifications [2, 31—37]. The currently used methods of cleaning CNT materials involve oxidative treatment with acid-oxidant mixtures, ozone [38], etc. They lead to partial or complete opening of the CNT ends and produce defects at the lateral surfaces.

The possible sites for sorption of gas impurity molecules in bundles of infinite, open and equal-diameter SWNTs are shown in Fig. 3. However, in practice such SWNT systems can have additional zones of impurity sorption. For example, nanotubes of different diameters form rather large channels parallel to the nanotube axes, which can be occupied by impurity molecules [39]. Besides, oxidation can produce interstices between the nanotubes in a bundle [18].

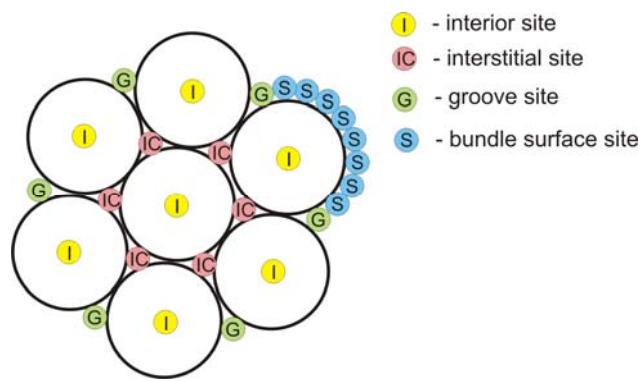

Fig. 3. Sites of possible sorption of gas impurity molecules in bundles of infinite, open and equal-diameter SWNTs.

We investigated Xe sorption in c-SWNT and o-SWNT powders at T = 78—200 K. The choice of the temperature interval and the impurity was dictated by the following considerations. The interaction of gas impurities with different parts of the CNT surface is most evident at low temperatures. Owing to their geometric configuration, SWNT bundles ideally (Fig. 3) have favorable sites where sorption of impurity molecules is energy-advantageous. A number of theoretical models were proposed [39—43] to describe the physical sorption and dynamics of admixed gas molecules at the surface and in the interstitial channels of SWNT bundles. According to mathematical simulations [43], the inner CNT surfaces and the interstices between the neighboring tubes at the surface of SWNT bundles (the grooves — G, Fig. 3) are the most energy-advantageous sites for sorbing impurity gas molecules. Xe was used because the SWNT—Xe system is already a well-studied "model" system [44—47]. A Xe atom is too large to penetrate into the interstitial channels (IC) of close-packed bundles of identical nanotubes whose energy of binding to impurity molecules is comparable to that at the inner surface [20]. Therefore, the Xe impurity is sorbed inside a nanotube (I), in a groove between two neighboring tubes at the outer surface of a bundle (G) and at the surface of the individual tubes forming the outer surface of a bundle (S) (see Fig. 3).

To obtain the necessary information about gas impurity desorption from CNT materials, a laboratory test bench (Fig. 4) was constructed for investigating the process of Xe sorption and desorption in a CNT powder at T = 78—200 K.

The measuring cell $V_1$ containing a CNT sample was filled with Xe at 12 torr and cooled slowly to T = 78 K. At this temperature the xenon available in the cell was sorbed by the CNT powder and condensed on the cell walls. The cell temperature was then increased in steps of 5 K. The Xe evaporated from the cell surface and was desorbed from different sites of the SWNT bundle surface. The evaporated Xe was condensed in the vessel $V_2$ cooled with liquid nitrogen. When the stepwise heating brought the pressure in the $V_1$—$V_2$ system to a constant value, the cell $V_1$ with the sample was separated from the vessel $V_2$. The Xe condensed in the vessel $V_2$ was evaporated and its pressure in the system was measured with the capacitive pressure transducer 5. With the volume of the system known, we could estimate the quantity of Xe desorbed from the sample at a particular temperature. To reduce the error due to the temperature gradient over the vessel $V_2$, the vessels $V_2$ and $V_3$ were minimized to the form of capillaries 1 mm in diameter. After each measurement run, Xe was recondensed from vessel $V_2$ to vessel $V_3$.

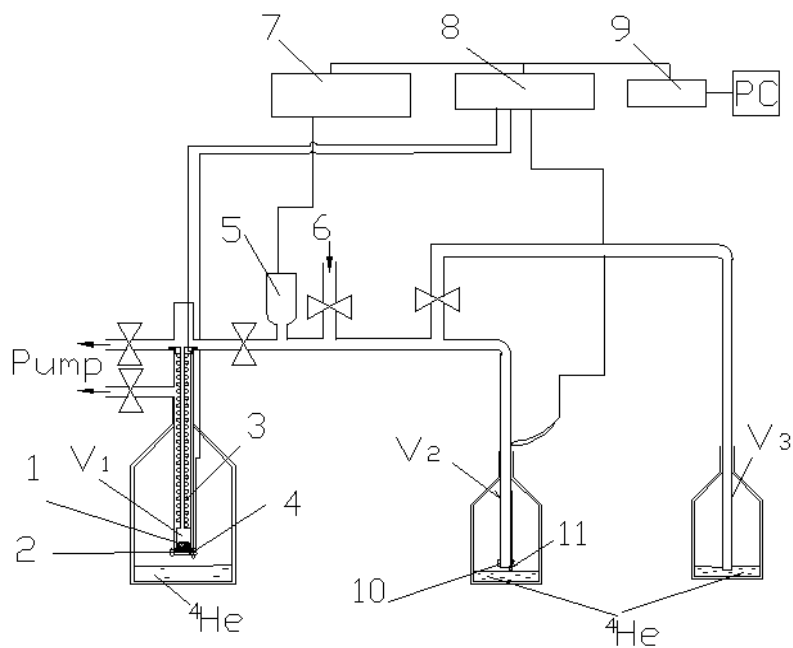

Fig. 4. Schematic view of the laboratory test bench for investigation of gas sorption-desorption in CNT samples at low temperatures.
    1 – Sample of nanotubes
    2, 3, 10 – Heaters
    4, 11 – Temperature sensor (silicone diode DT – 470)
    5 – Pressure transducer (capacitance manometer MKS Baratron 627B)
    6 – Gas input
    7 – Digital multimeter (Keithley 2700)
    8 – Temperature controller (Cryo-Con model 34)
    9 – Matching device (Advantech PCI – 1670)

## *2.2. Results and discussion*

The sorption properties of the starting pure c-SWNT powder (0.0416 g) were investigated through thermo-programmed desorption (see above). Fig. 5 illustrates the temperature distribution of the desorbed impurity. The greatest quantities of Xe were desorbed at $T = 125$—$135$ K. In the case of close-packed bundles of infinite equal-diameter SWNTs (Fig. 3), the highest desorption of Xe in this temperature interval is expected from the grooves at the outer bundle surface (G) and from the interior channels of some nanotubes (I) because Xe atoms have the highest and nearly equal binding energies at these sites [47]. In our powder the desorption can be enhanced considerably by removal of Xe atoms from the axial large-diameter channels (IC). Such channels are possible in bundles of nanotubes of varying diameters. Xenon can penetrate into interior channels through defects at the ends or the lateral surfaces that can be present in some nanotubes of the starting powder. A rather small quantity of Xe was also desorbed at $T = 100$—$105$ K, which may be due to removal of the layers (S) of Xe molecules that form at the outer surface of SWNT bundles.

To open the nanotube ends, a portion (0.0705 g) of the starting powder was placed into a capsule which was then evacuated for 8 hours and heated to 450 °C. At this temperature the capsule was filled with air for 12 min. under atmospheric pressure. According to the literature data, the ends of over 90% of CNTs can be opened through this procedure [48]. Thereafter, the capsule was evacuated again to about $10^{-3}$ mm Hg, heated to 750 °C and held at this temperature for an hour to remove the gaseous oxidation products. The post-treatment weighting showed a loss of ~ 5% in the powder mass.

The sorption properties of the nanotubes with the opened ends were then investigated using the same thermoprogrammed desorption technique (see above). The oxidation-induced opening of the CNT ends made the inner CNT surfaces and the intertube interstice in the bundles accessible to Xe sorption [18], which enhanced the sorption capacity of the SWNT powder almost fivefold as compared to the starting material (see Fig. 5).

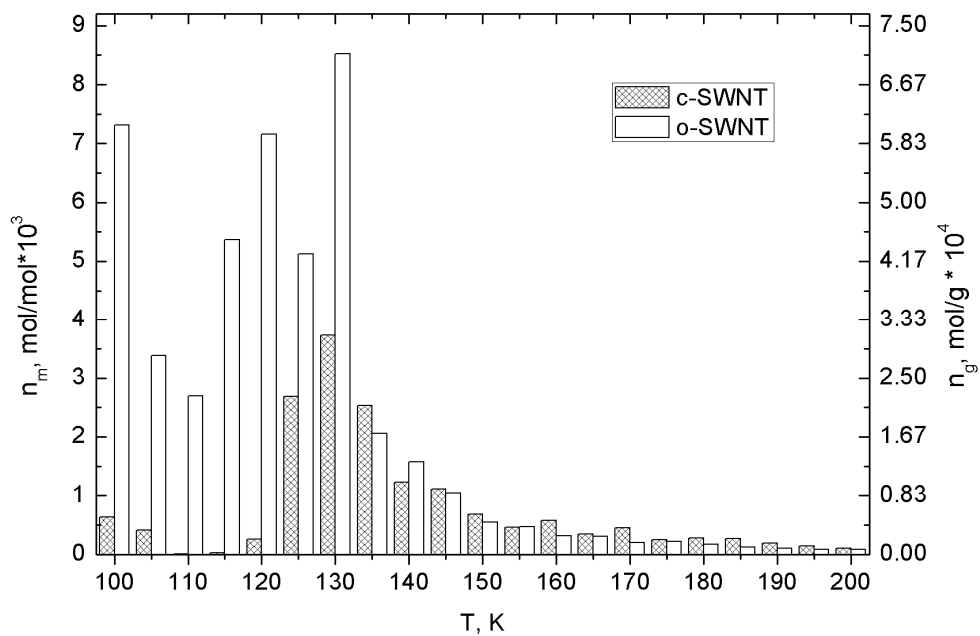

Fig. 5. Temperature distribution of Xe impurity (mole per mole and mole per gram) desorbed from powder samples of c-SWNTs (dark columns) and o-SWNTs (light columns).

## 3. Radial thermal expansion of xenon-saturated single-walled carbon nanotubes. Discussion

The radial thermal expansion of Xe-saturated SWNTs was also investigated on the compacted sample used previously to measure the LTECs of pure SWNTs. The measurement technique is described in Section 1. Immediately before measurement, the cell with a pure CNT sample was evacuated at room temperature for 96 hours and then filled with Xe at 760 mm Hg. The evacuated measuring cell of the dilatometer with the sample in the Xe atmosphere was cooled to 90 K. At this temperature it was evacuated again and then cooled to liquid helium temperature. The thermal expansion was measured in vacuum down to $1 \cdot 10^{-5}$ mm Hg.

The temperature dependence of the LTEC taken on a Xe—SWNT sample in the interval 2.2—75 K is shown in Fig. 6 (curve 1). The sharp increase in the LTECs of the Xe-saturated sample (cf. curves 1, 3) can reasonably be attributed to the heavy Xe atoms affecting the transverse vibrations of the nanotubes in the direction perpendicular to their surface. At low temperatures the Grüneisen coefficients of such vibrations are negative in two-dimensional (graphene) or quasi-two-dimensional (graphite, nanotubes) carbon systems [8, 49] and positive in a three-dimensional carbon modification (diamond). The formation of SWNT bundles and the sorption of impurity atoms at the surface or inside the nanotubes generate three-dimensional features in the system. As a result, the negative Grüneisen coefficients of such system decrease in magnitude or become positive. The thermal expansion coefficients are expected to behave in a similar way. That is why the negative contribution to the radial thermal expansion of Xe-saturated SWNT bundles decreases and shifts towards lower temperatures (see Fig. 6; cf. curves 1, 2 and 3).

In contrast to pure CNTs the thermal expansion of Xe-saturated SWNTs is similar during the first heating and in the subsequent heating and cooling runs. It is possible that the first heating of pure SWNT bundles with xenon can make the system more rigid and its geometry insensitive to heating at low temperatures.

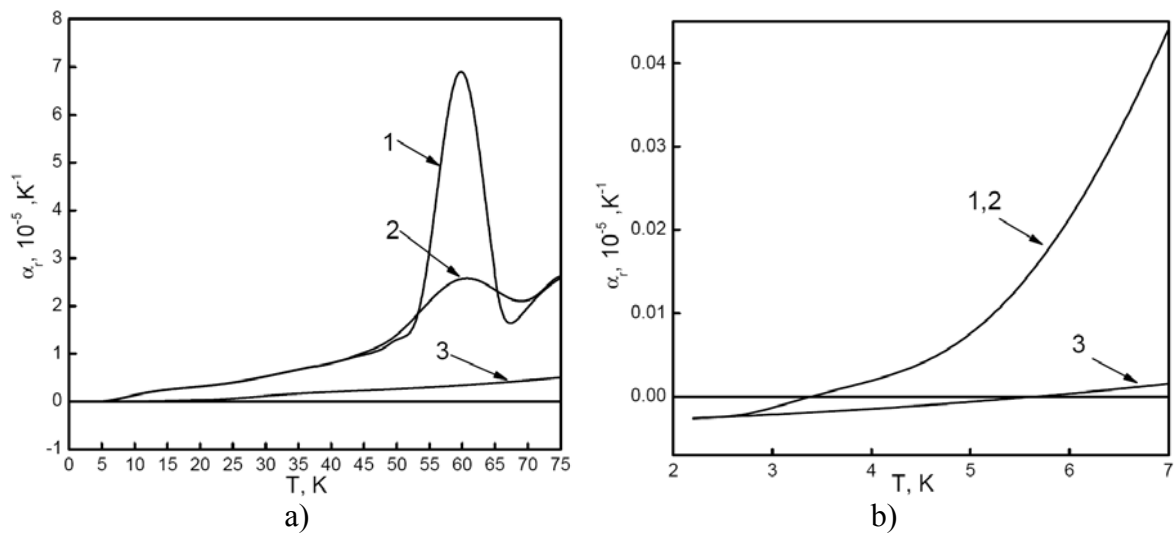

Fig. 6. The radial LTECs of SWNT bundles: 1 – Xe-saturated bundles, 2 – after partial Xe removal at T = 110 K, 3 – pure SWNTs ($\alpha_r$) at T = 2.2—75 K (a) and 2.2—7 K (b), compacted sample.

It is interesting that the LTECs have maximum values in the interval 50—65 K, which may be a manifestation of spatial redistribution of the Xe atoms in the SWNT bundles. The simulation (by Wang-Landau algorithm) [23, 50] of the potential energy for a system of SWNT bundles saturated with inert gases predicted peaks in the temperature dependence of the heat capacity at T = 50—100 K, attributed to reordering of the impurity atoms.

To test the prediction, it was necessary to remove the Xe impurity from the surface of the SWNT bundles. For this purpose, the sample was heated to T = 110 K. This temperature causes intensive desorption of Xe from the sample surface but leaves it undisturbed in the grooves of the SWNT bundles (G) and the inner interstices (I) of the nanotubes having surface defects (Fig. 5). The sample was kept at T = 110 K until the desorbed Xe was entirely removed and the pressure in the measuring cell reached $\sim 1\cdot 10^{-5}$ mm Hg. The sample was then cooled to T = 2.2 K and the thermal expansion was measured again (Fig. 6, curve 2). It is seen that the LTEC peak is much lower after Xe was removed from the SWNT bundle surfaces. However, this partial Xe desorption leaves the temperature dependence of the LTEC practically unaffected outside the interval of the peak. This suggests that the Xe atoms residing on the bundle surface have only a small effect upon the thermal expansion of SWNT bundles when the process of spatial redistribution of atoms are absent.

## Conclusions

This is the first time that the temperature dependences of the radial thermal expansion coefficients $\alpha_r$ (T) of pure and Xe-saturated SWNT bundles have been investigated experimentally at low temperatures. The measurements were made on heating and cooling the samples in the interval 2.2—120 K using a capacitance dilatometer.

The dependence $\alpha_r$ (T) measured on the first heating showed very strong nonequilibrium effects, and in the interval 3.2—120 K it differed significantly from the well reproducible equilibrium dependences $\alpha_r$ (T) that were found on subsequent heating and cooling runs in this measurement.

The equilibrium coefficients of the radial thermal expansion $\alpha_r$ (Fig. 2, curve 1) are positive above 5.5 K and negative at lower temperatures. The nonequilibrium coefficients of the radial thermal expansion $\alpha_r$ (Fig. 2, curve 2) are negative in the interval 2.2—82 K. It is assumed that the non-equilibrium $\alpha_r$-values measured on the first heating of the sample are due to the irreversible alignment and ordering of the bundle positions and the nanotubes in the bundles at rising

4temperature. As this occurs, the density of the system increases, and the thermal expansion becomes negative.

The qualitative interpretation of the equilibrium dependence $\alpha_r$ (T) was based on the theoretical conclusions about the Grüneisen coefficients for carbon modifications [8]. The Grüneisen coefficient and the radial thermal expansion of nanotubes are negative at reasonably low temperatures [8], which is determined mainly by the contribution of the transverse acoustic vibrations perpendicular to the nanotube surfaces. However, in the experiment the temperature interval of the negative thermal expansion is much narrower in comparison with the theoretical predictions. This is most likely because the cited theory [8] investigated individual nanotubes. Additional contributions to the thermal expansion come into play in SWNT bundles. First, there is a positive contribution $\alpha_g$, generated by the variations of the intertube gaps with temperature. In addition, the nanotube interaction in the bundles suppresses the negative contribution from the transverse acoustic vibrations perpendicular to the nanotube surfaces [8]. These two positive contributions to the thermal expansion of the SWNT bundles decrease both the magnitude and the temperature interval of the negative thermal expansion.

The saturation of SWNT bundles with xenon brings about new features in their thermal expansion.

1) The magnitude of $\alpha_r$ increases sharply in the whole range of temperature investigated. This is because the Xe impurity suppresses the negative contribution to the thermal expansion from the transverse acoustic vibrations perpendicular to the nanotube surfaces [8].

2) The dependence $\alpha_r$ (T) has a peak in the interval 50—65 K, which appears to be due to the spatial redistribution of the Xe atoms over the SWNT bundle surfaces. Removal of the Xe impurity from these surfaces decreases the peak significantly but leaves the temperature dependence of the LTEC practically unchanged outside the interval of the peak. This suggests that the Xe atoms atoms located at the bundle surfaces have little effect on the thermal expansion of SWNT bundles when the processes of their spatial redistribution are inoperative.

3) For the Xe saturated material there is no non-equilibrium thermal expansion behaviour such as was observed during the first heating of the sample and attributed to irreversible alignment and ordering of the bundle positions and the nanotubes in the bundles at rising temperature. It is likely that the saturation with Xe makes the system of SWNT bundles more rigid and its geometry insensitive to heating in a low temperature interval.

Finally, the employed technique of thermoprogrammed desorption has also enabled us to measure the temperature dependence of Xe desorption from both open and closed SWNT bundles.

We wish to thank Prof. V.M. Loktev for valuable discussion.

The authors are indebted to the Science and Technology Center of Ukraine (STCU) for the financial support of this study (project No 4266).## Bibliography

1. S. Iijima, *Nature* **354**, 56 (1991)
2. A. V. Eletskii, *Phys. Usp.* **47**, 1119 (2004)
3. H. Jiang, B. Liu and Y. Huang, *J. Eng. Mater. Technol.* **126**, 265 (2004)
4. Y. Kwon, S. Berber and D. Tomanek, *Phys. Rev. Lett.* **92**, 015901 (2004)
5. N. M. Prakash. *Determination of coefficient of thermal expansion of single-walled carbon nanotubes using molecular dynamics simulation*. The Florida State University. Dissertation Master of Science, (2005) p. 54.
6. C. Li and T. Chou, *Phys. Rev. B* **71**, 235414 (2005)